\documentclass[aps,twocolumn,showpacs,preprintnumbers,amsmath,amssymb,floatfix]{revtex4-2}

\usepackage{graphicx}
\usepackage{longtable}
\usepackage{dcolumn}
\usepackage{bm}
\usepackage{color}
\usepackage[normalem]{ulem}
\usepackage[colorlinks=true, urlcolor=blue, linkcolor=blue, citecolor=blue]{hyperref}
\usepackage[all]{hypcap}
\newcommand{\beq}{\begin{equation}}
\newcommand{\eeq}{\end{equation}}
\newcommand{\bea}{\begin{eqnarray}}
\newcommand{\eea}{\end{eqnarray}}

\begin{document}

\title{Spin gapped metals: A novel class of materials for multifunctional spintronic devices}

\author{E. \c{S}a\c{s}{\i}o\u{g}lu$^1$}\email{ersoy.sasioglu@physik.uni-halle.de}
\author{M.  Tas$^2$}
\author{S. Ghosh$^{3}$}
\author{W. Beida$^{4}$}
\author{B. Sanyal$^{3}$}
\author{S. Bl\"{u}gel$^{4}$}
\author{I. Mertig$^{1}$}
\author{I. Galanakis$^{5}$}\email{galanakis@upatras.gr}
\affiliation{$^{1}$Institute of Physics, Martin Luther University Halle-Wittenberg, 06120 Halle (Saale), 
Germany \\
$^{2}$Department of Physics, Gebze Technical University, 41400 Kocaeli, Turkey \\
$^{3}$Department of Physics and Astronomy, Uppsala University, 75120 Uppsala, Sweden \\
$^{4}$Peter Gr\"unberg Institut, Forschungszentrum J\"ulich 
and JARA, 52425 J\"ulich, Germany \\
$^{5}$Department of Materials Science, School of Natural Sciences, University of Patras, GR-26504 Patras, 
Greece}

\date{\today}

\begin{abstract}
%% Text of abstract
Gapped metals, a recently proposed class of materials, possess a band gap slightly above 
or below the Fermi level, behaving as intrinsic p- or n-type semiconductors without requiring 
external doping. Inspired by this concept, we propose a novel material class: "spin gapped metals". 
These materials exhibit intrinsic p- or n-type character independently for each spin channel, 
similar to dilute magnetic semiconductors but without the need for transition metal doping.
A key advantage of spin gapped metals lies in the absence of band tails that exist within 
the band gap of conventional p- and n-type semiconductors. Band tails degrade the performance 
of devices like tunnel field-effect transistors (causing high subthreshold slopes) and negative 
differential resistance tunnel diodes (resulting in low peak-to-valley current ratios).
Here, we demonstrate the viability of spin gapped metals using first-principles electronic band 
structure calculations on half-Heusler compounds. Our analysis reveals compounds displaying both
gapped metal and spin gapped metal behavior, paving the way for next-generation multifunctional 
devices in spintronics and nanoelectronics.
\end{abstract}

\maketitle

%% main text
\section{Introduction}

\begin{figure*}
\centering
\includegraphics[width=0.9\textwidth]{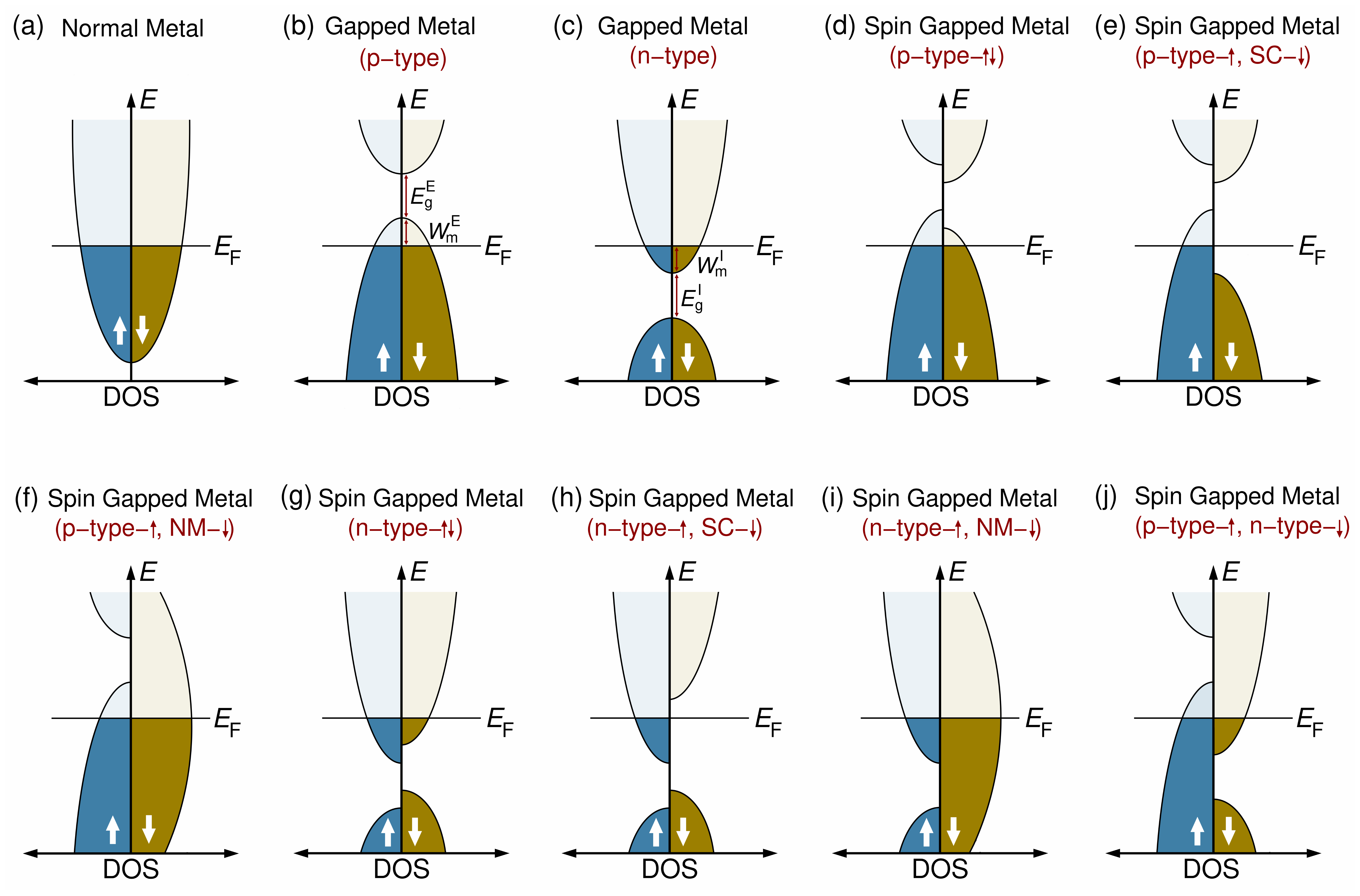}
\vspace{-0.2 cm}
\caption{Schematic representation of the density of states (DOS) of a 
normal metal (a), gapped metals (b-c), and Spin gapped metals (d-j). 
The arrows represent the two possible spin directions. The horizontal line 
depicts the Fermi level $E_\mathrm{F}$. NM stands for normal-metal and SC 
for semiconductor.}
\label{fig1}
\end{figure*}

An essential challenge in materials science involves uncovering materials 
possessing unique characteristics that can bolster device performance. 
Thermoelectricity, which involves converting heat into electricity via the 
Seebeck effect, stands as a pivotal phenomenon in contemporary technology, 
enabling the harnessing of waste heat from both industrial and household
processes \cite{Bell2008,Champier2017}. While doped semiconductors are 
traditionally favored for thermoelectric applications, recent suggestions 
propose a new category of metals termed "gapped metals" as potential 
substitutes for doped semiconductors \cite{Ricci2020,Malyi2020,Khan2023}. 
As illustrated in Figure\,\ref{fig1}(a-c), the density of states (DOS) for 
these gapped metals reveal a semiconductor-like gap, which distinguishes 
them from conventional metals. Specifically, the Fermi level intersects 
either the valence band, resulting in an excess of holes (p-type gapped 
metals), or the conduction band, leading to an excess of electrons (n-type 
gapped metals). A key advantage of gapped metals lies in the absence of band 
tails, which originates from the strong doping and doping fluctuations in 
conventional p-/n-type semiconductors and they have been studied in detail 
by many authors using different approaches 
\cite{efros1974density,chakraborty2001density,kane1985band,van1992theory,sant2017effect,bizindavyi2018band,schenk2020tunneling}. 
Band tails degrade the performance of devices like tunnel field-effect transistors 
(FETs) with increased subthreshold slope (SS) and negative differential resistance (NDR) 
tunnel diodes with lowered peak-to-valley current ratio.

Gapped metals, a subclass of "cold metals", share many similar characteristics \cite{wang2023cold}. 
Cold metals  have garnered significant interest for their potential in next-generation electronics, 
particularly devices like NDR tunnel diodes \cite{sasioglu2023theoretical} and steep-slope FETs.
When used as source and drain electrodes in FETs, gapped metals, similar to cold metals, 
can filter the transmission of high-energy electrons in the subthreshold region, leading 
to sub-60 mV/dec SS values and reduced leakage current in the off-state 
\cite{qiu2018dirac,liu2020switching,marin2020lateral,logoteta2020cold,tang2021steep}. 
Furthermore, gapped metals, like cold metals, can be employed in metal-semiconductor Schottky 
barrier diodes instead of the normal metals to surpass the thermionic emission limit, typically 
characterized by an ideality factor ($\eta$) of 1 at room temperature. These advancements pave 
the way for the development of low-power nanoelectronic circuits \cite{shin2022steep}.

In this Article, we introduce the term 'spin gapped metals', extending the concept 
of gapped metals to encompass magnetic materials. This terminology encapsulates the 
distinct spin-dependent electronic band structures of these materials. Such an extension 
would result in a diverse range of behaviors stemming from the distinct characteristics 
of each metal. Seven distinct scenarios are presented in Figure~\ref{fig1}(d-j) to 
illustrate this concept. By combining n- or p-type gapped metallic behavior for the 
spin-up electronic band structure with various behaviors such as normal metallic, typical 
semiconducting, or n(p)-type gapped metallic behavior in the spin-down electronic 
band structure, a broader range of implications for device applications could 
be achieved.  Analogous to how gapped metals are seen as counterparts to doped 
semiconductors, one might view spin gapped metals as counterparts to diluted 
magnetic semiconductors \cite{Sato2010,Lei2022,Tu2014,Kroth2006}, such as Mn-doped 
GaAs, without requiring the incorporation of magnetic atoms through doping.

\section{Results and discussion}

Gapped metals and spin gapped metals can be qualitatively described by four key 
electronic structure parameters as shown on the schematic density of states 
in Figure~\ref{fig1}. These parameters include the internal band gap ($E_{\mathrm{g}}^{\mathrm{I}}$), 
the external band gap ($E_{\mathrm{g}}^{\mathrm{E}}$) and internal (external) metallic 
bandwidth $W_{\mathrm{m}}^{\mathrm{I}}$ ($W_{\mathrm{m}}^{\mathrm{E}}$). $W_{\mathrm{m}}^{\mathrm{E}}$ 
represents the energy difference between the Fermi level and the valence band maximum for 
p-type gapped and spin gapped metals, while $W_{\mathrm{m}}^{\mathrm{I}}$ represents 
the energy difference between the conduction band minimum and the Fermi level for 
n-type materials. The importance of W$_{\mathrm{m}}^{\mathrm{E}}$  is evident in field-effect 
transistors (FETs) when gapped metals are used as source electrodes, as demonstrated 
in reference\,\cite{liu2020switching}  for p-type NbTe$_2$-based transistors. A smaller 
$W_{\mathrm{m}}^{\mathrm{E}}$ value translates to a lower SS for the transistor, leading 
to reduced energy consumption. FETs designed with both source and drain electrodes composed 
of p-type gapped metals (or cold metals) are predicted to exhibit not only transistor functionality 
but also the gate-voltage tunable NDR effect \cite{yin2022computational}. The spin degree of 
freedom in spin gapped metals opens exciting possibilities for the development of multifunctional 
devices, which we will comment on later in this paper.

The recently proposed concept of "gapped metals" \cite{Malyi2020}, although new, has 
already been observed in many materials reported earlier \cite{sakuma2013electronic,bigi2020direct,may2009influence,huang2015new}. 
This newfound understanding has ignited renewed interest in identifying materials that 
exemplify this promising class \cite{Khan2023,Ricci2020}. Similarly, materials exhibiting 
spin gapped metal-like behavior might exist in previously published works, but haven't 
been categorized as such due to the lack of a suitable framework. In this work, we 
address this gap by extending the concept of gapped metals to magnetic materials, 
introducing the term "spin gapped metals". This framework allows for the classification 
of new materials based on their spin-dependent electronic structures. Motivated 
by our recent discovery of spin-polarized two-dimensional electron/hole gas at the 
interfaces of semiconducting 18-valence electron half-Heusler compounds (also known as semi-Heusler compounds), 
which displayed spin gapped metal-like interface states \cite{Emel2023}, we aim to explore 17-valence 
electron and 19-valence electron half-Heusler compounds as potential candidates for exhibiting 
spin gapped metal behavior. The so-called 18-valence electrons half-Heusler 
compounds, such as CoTiSb, FeVSb, or NiTiSn, are renowned semiconductors exhibiting 
remarkably high-temperature thermoelectric properties \cite{Ma2017,Jung2000,
PIERRE1994,Tobola2000,Ouardi2012,Mokhtari2018,Emel2023,Emel2023b,Xu2013}. Seeking 
spin gapped metals within the half-Heusler compounds, deviating from 
the 18-valence electrons semiconducting Heusler compounds seems a logical step.

\begin{table*}[t]
\caption{\label{table1}
Lattice constants $a_0$, valence electron number Z$_T$, sublattice and total magnetic 
moments, spin polarization at the Fermi level (see text for definition), spin gap type 
per spin direction, the distance of the Fermi level from the edge of the band which it 
crosses $W_{\mathrm{m}}^{\mathrm{I,E}(\uparrow/\downarrow)}$ (see text for more details), 
energy gap per spin direction  $E_{\mathrm{g}}^{\mathrm{I,E}(\uparrow/\downarrow)}$, 
formation energy ($E_\text{form}$), and convex hull distance energy ($\Delta E_\text{con}$) 
for the compounds under study. The $a_0$, $E_\text{form}$ and $\Delta E_\text{con}$ values 
are taken from the Open Quantum Materials Database~\cite{oqmd,Saal2013,Kirklin2015}.} 
\begin{tabular}{lcccccllllll}
\hline
\hline
Compound & $a_0$ & Z$_T$ & m$_X$ & m$_{Y}$ & m$_\text{total}$ & SP & Spin gap type & $W_{\mathrm{m}}^{\mathrm{I,E}(\uparrow/\downarrow)}$ & $E_{\mathrm{g}}^{\mathrm{I,E}(\uparrow/\downarrow)}$ & $E_\text{form}$ &  $E_\text{con}$  \\
   $XYZ$   & ({\AA}) & & ($\mu_B$) & ($\mu_B$) &  ($\mu_B$)   & (\%) &            & (eV) & (eV) & (eV/at.) & (eV/at.)  \\ \hline
\multicolumn{12}{c}{Spin gapped metals} \\
FeTiSb & 5.94 & 17 & -1.45 & 0.53  &  -0.95 & 68  & p-type-$\uparrow$/p-type-$\downarrow$  & 0.58/0.12  & 0.47/0.76   & -0.39 & 0.04   \\
FeZrSb & 6.15 & 17 & -1.34 & 0.34  &  -1.00 & 100 & p-type-$\uparrow$/SC-$\downarrow$      & 0.64/      & 0.90/       & -0.45 & 0.03   \\
FeHfSb & 6.11 & 17 & -1.26 & 0.27  &  -1.00 & 100 & p-type-$\uparrow$/SC-$\downarrow$      & 0.60/      & 1.28/       & -0.39 & 0.00   \\
FeVSn  & 5.87 & 17 & -1.85 & 1.03  &  -0.88 & 36  & NM-$\uparrow$/p-type-$\downarrow$ & {\hspace{0.6cm}}/0.15  & {\hspace{0.6cm}}/0.59 & 0.06 & 0.18  \\
FeNbSn & 6.00 & 17 & -1.38 & 0.40  &  -1.00 & 100 & p-type-$\uparrow$/SC-$\downarrow$     & 0.70/ & 0.16/  & -0.10    & 0.06    \\
FeTaSn & 5.99 & 17 & -1.29 & 0.32  &  -1.00 & 100 & p-type-$\uparrow$/SC-$\downarrow$      & 0.70/ & 0.56/  & -0.06    & 0.09   \\
CoTiSn & 5.93 & 17 & -0.42 & -0.45 &  -0.94 & 74  & p-type-$\uparrow$/p-type-$\downarrow$  & 0.37/0.06 & 0.94/0.84   & -0.38  & 0.07    \\
CoZrSn & 6.15 & 17 & -0.67 & -0.18 &  -0.95 & 81  & p-type-$\uparrow$/p-type-$\downarrow$  & 0.41/0.09 & 0.97/0.74   & -0.46  & 0.05    \\
CoHfSn & 6.11 & 17 & -0.55 & -0.15 &  -0.79 & 66  & p-type-$\uparrow$/p-type-$\downarrow$  & 0.38/0.24 & 1.02/0.71   & -0.43 & 0.04     \\
RhTiSn & 6.17 & 17 & -0.07 & -0.73 &  -0.87 & 70  & p-type-$\uparrow$/p-type-$\downarrow$  & 0.56/0.09 & 0.63/0.65   & -0.60 & 0.06   \\
IrTiSn & 6.20 & 17 & -0.08 & -0.61 &  -0.76 & 33  & p-type-$\uparrow$/p-type-$\downarrow$  & 0.45/0.13 & 0.71/0.67   & -0.61 & 0.02   \\
NiZrIn & 6.22 & 17 & -0.11 & -0.31 &  -0.55 & 53  & p-type-$\uparrow$/p-type-$\downarrow$  & 0.49/0.34 & 0.30/0.30   & -0.31 & 0.12     \\
PtTiIn & 6.24 & 17 & -0.04 & -0.84 &  -1.00 & 100 & p-type-$\uparrow$/SC-$\downarrow$      & 0.45/     & 0.78/       & -0.56  & 0.10     \\
CoVSb  & 5.81 & 19 & -0.33 & 1.41  &  1.00  & 100 & n-type-$\uparrow$/SC-$\downarrow$      & 0.79/     & 0.35/        & -0.19  & 0.07    \\
RhVSb  & 6.07 & 19 & -0.21 & 1.33  &  1.00  & 100 & n-type-$\uparrow$/SC-$\downarrow$      & 0.87/ & 0.25/  & -0.31  & 0.10       \\
PdTiSb & 6.24 & 19 & -0.04 & 0.98  &  0.89  & 87  & n-type-$\uparrow$/n-type-$\downarrow$  & 0.59/0.19 & 0.49/0.41 & -0.51 & 0.09   \\
PtTiSb & 6.26 & 19 & -0.05 & 1.05  &  0.99  & 93  & n-type-$\uparrow$/n-type-$\downarrow$  & 0.53/0.02 & 0.74/0.78 & -0.65 & 0.11   \\
NiVSn  & 5.87 & 19 & -0.03 & 1.15  &  1.00  & 100 & n-type-$\uparrow$/SC-$\downarrow$      & 0.41/      & 0.67/     & -0.08 & 0.15   \\
NiVSb  & 5.88 & 20 & 0.03 & 2.12   &  2.00  & 100 & n-type-$\uparrow$/SC-$\downarrow$      & 0.68/     & 0.80/     & -0.12 & 0.13    \\
CuVSb  & 6.06 & 21 & 0.04 & 3.04   &  2.95  &  40 & n-type-$\uparrow$/NM-$\downarrow$      & 1.44/      &  0.15/   &  {\hspace{0.15cm}}0.21 & 0.24   \\
\multicolumn{12}{c}{Gapped metals} \\
NiHfIn & 6.16 & 17 &  0.00 & 0.00 &  0.00 & 0 & p-type-$\uparrow$/p-type-$\downarrow$ & 0.45/0.45 & 0.32/0.32 & -0.26 &  0.17    \\
NiTiSb & 5.93 & 19 &  0.00 & 0.00 &  0.00 & 0 & n-type-$\uparrow$/n-type-$\downarrow$ & 0.69/0.69 & 0.61/0.61 & -0.52 & 0.05     \\
NiZrSb & 6.15 & 19 &  0.00 & 0.00 &  0.00 & 0 & n-type-$\uparrow$/n-type-$\downarrow$ & 0.88/0.88 & 0.75/0.75 & -0.62 & 0.03     \\
NiHfSb & 6.10 & 19 &  0.00 & 0.00 &  0.00 & 0 & n-type-$\uparrow$/n-type-$\downarrow$ & 0.96/0.96 & 0.67/0.67 & -0.53 & 0.06     \\
NiNbSn & 6.00 & 19 & 0.00  & 0.00 &  0.00 & 0 & n-type-$\uparrow$/n-type-$\downarrow$ & 0.73/0.73 & 0.88/0.88 & -0.19 & 0.09     \\ 
\hline
\hline
\end{tabular}
\end{table*}

The choice to search for spin gapped metals among the Heusler compounds  is natural since 
the latter constitute a vast family of intermetallic compounds, currently comprising 
over 2000 members \cite{Graf2011,Tavares2023,Chatterjee2022}. Within this family, numerous 
novel behaviors have been both experimentally and computationally identified, 
rendering them appealing for a wide array of device applications. 
These behaviors include half-metallicity \cite{Galanakis2023}, spin gapless 
semiconducting behavior \cite{Ouardi2013,Gao2019,Xu2013}, spin-filtering \cite{Galanakis2016}, 
and more. Their adaptability regarding element substitution drives ongoing 
research in this field, and extended databases were built using the first-principles 
calculations resulting in the prediction of hundreds of new Heusler compounds which 
were later grown experimentally \cite{Galanakis2023,Gao2019,Ma2017,Gillessen2009,
Gillessen2010,Faleev2017a,Faleev2017b,Faleev2017c,Marathe2023,Sanvito2017}.

To pursue the objective of identifying spin gapped metals, we conducted a search in 
the Open Quantum Materials Database (OQMD) \cite{oqmd} and identified 25 half-Heusler compounds, 
as outlined in Table\,\ref{table1}. These compounds typically possess 17 or 19 valence 
electrons per unit cell, though some exceptions with 20 or 21 valence 
electrons exist. Our initial criterion for selecting compounds for the study was 
their formation energy, $E_\text{form}$, which we aimed to be negative, with a 
few exceptions like FeVSn and CuVSb included for completeness, as their values 
were close to zero despite being positive. However, negative $E_\text{form}$ alone 
does not guarantee stability. The convex hull distance $\Delta E_\text{con}$, 
which represents the energy difference between the studied structure and the most 
stable phase or a mixture of phases, is also crucial. Typically, values less than 
0.2 eV/atom is desired to facilitate the growth of a material in a metastable 
form, such as a thin film or a nanostructure. All materials selected from OQMD for 
our study exhibit  $\Delta E_\text{con}$ values less than the cutoff of 0.2 eV/atom, 
except for CuVSb, which has a value of 0.24 eV/atom. The formation and convex hull 
distance energies from OQMD are detailed in Table \ref{table1}.

The bulk half-Heusler compounds $XYZ$ that we consider in this work crystallize in the 
cubic $C1_\mathrm{b}$ lattice structures (see figure 2 in reference\,\cite{Galanakis2023}). 
The space group is the $F\overline{4}3m$ and actually consists of four interpenetrating 
f.c.c. sublattices; one is empty and the other three are occupied by the $X$, $Y$, and 
$Z$ atoms. The unit cell is an f.c.c. one with three atoms as a basis along the long 
diagonal of the cube: $X$ at $(0\:0\:0)$, $Y$ at $(\frac{1}{4}\:\frac{1}{4}\:\frac{1}{4})$ 
and $Z$ at $(\frac{3}{4}\:\frac{3}{4}\:\frac{3}{4})$ in Wyckoff coordinates. We adopted 
the lattice constants calculated in the Open Quantum Materials Database (OQMD) for all 
twenty-five materials \cite{oqmd,Saal2013,Kirklin2015}, and we present them in the  second column in Table\,\ref{table1}. 
Our tests show that the lattice constants presented in OQMD, where the Perdew-Burke-Ernzerhof 
(PBE) functional has been used for the exchange-correlation potential \cite{Perdew1996}, 
differ less than 1 \%\ from the PBE equilibrium ones calculated with the electronic band 
structure method employed in the current study.

The ground-state first-principles electronic band-structure calculations for all studied 
compounds were carried out using the \textsc{QuantumATK} software package \cite{QuantumATK,QuantumATKb}. 
We use linear combinations of atomic orbitals (LCAO) as a basis set together with norm-conserving 
PseudoDojo pseudopotentials \cite{VanSetten2018}. The PBE parameterization to the 
generalized-gradient-approximation of the exchange-correlation potential is employed \cite{Perdew1996}. 
Note that due to the metallic character of the compounds under study, the GGA provides a more 
accurate description of the ground state properties concerning more complex hybrid functionals
\cite{Emel2023b,Meinert2013}. For determination of the ground-state properties of the bulk compounds, 
we use a $16 \times 16 \times 16$ Monkhorst-Pack $\mathbf{k}$-point grid \cite{Monkhorst1976}. 

Our calculations do not account for spin-orbit coupling. This 
interaction is anticipated to introduce states within the 
energy gap, particularly in materials exhibiting half-metallic 
ferromagnetism. These materials are characterized by a 
minority-spin band gap encompassing the Fermi level, while the 
majority-spin band displays metallic behavior.
However, studies by Mavropoulos et al. \cite{Mavropoulos2004a} 
utilizing relativistic first-principles calculations have 
demonstrated that the influence of spin-orbit coupling is 
negligible in NiMnSb half-metallic Heusler compound. This 
finding was further generalized by Mavropoulos et al. 
\cite{Mavropoulos2004b} to encompass not only half-metals but 
also near-half-metallic materials such as PdMnSb and PtMnSb. 
These materials maintain nearly 100\%\ spin polarization at 
the center of the energy gap.
The authors in Ref. \cite{Mavropoulos2004b} explained this 
behavior using perturbation theory. They showed that 
DOS exhibits a quadratic dependence on the 
strength of spin-orbit coupling. Furthermore, the spin-down 
DOS weakly mirrors the spin-up DOS, resulting in a minimal 
impact of spin-orbit coupling.

Our search resulted in twenty-five half-Heusler compounds which fullfil the criteria 
discussed above and our calculations indicate that are either gapped or spin gapped metals. In 
Table\,\ref{table1} we have compiled our results. We have five compounds that are 
non-magnetic and belong to the gapped metals. Among them, only NiHfIn has 17 valence 
electrons per unit cell while the other four have 19 valence electrons. Since the 
18 valence electrons compounds are semiconductors, one would expect NiHfIn to be a 
p-type gapped metal with the Fermi level crossing the valence band and the rest 
n-type gapped metals with the Fermi level crossing the conduction band. Our 
calculated band structures confirm these predictions as shown in Table\,\ref{table1}. 
In Table\,\ref{table1} we also include the distance of the Fermi level from the edge 
of the band which it crosses, $W_{\mathrm{m}}^{\mathrm{I,E}(\uparrow/\downarrow)}$, 
The 
$W_{\mathrm{m}}^{\mathrm{I,E}(\uparrow/\downarrow)}$ value is followed by the 
$E_{\mathrm{g}}^{\mathrm{I,E}(\uparrow/\downarrow)}$ one which is the width of the 
energy gap. For each quantity, we provide two values separated by a slash 
corresponding to the two spin directions. As it is obvious for the gapped metals 
the values for both spin directions are identical. Both 
$W_{\mathrm{m}}^{\mathrm{I,E}(\uparrow/\downarrow)}$ and 
$E_{\mathrm{g}}^{\mathrm{I,E}(\uparrow/\downarrow)}$ values are less than 1 eV and 
are comparable.

\begin{figure*} [t]
\centering
\includegraphics[width=0.95\textwidth]{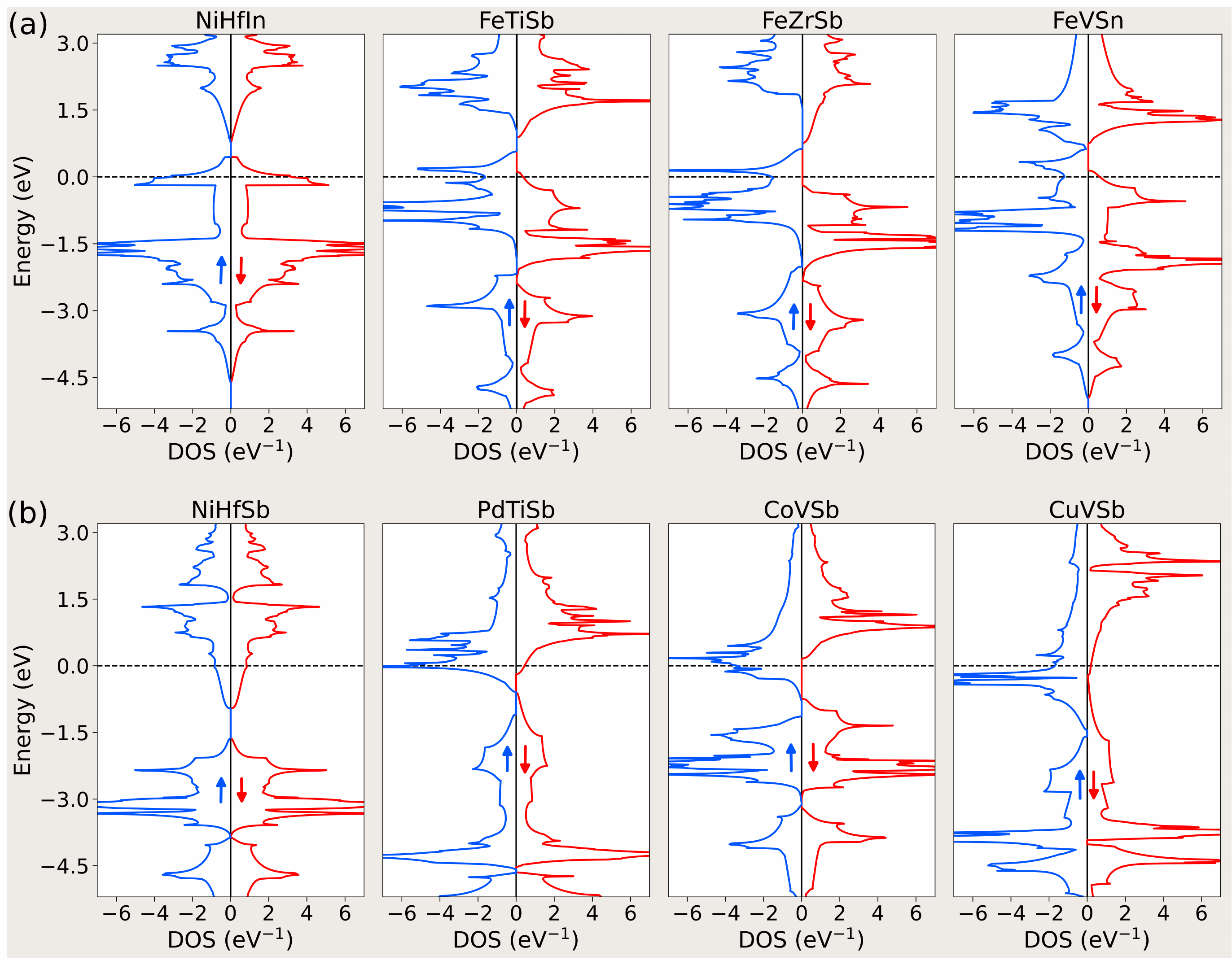}
\vspace{-0.2 cm}
\caption{Spin-resolved total density of states for selected gapped and spin gapped half-Heusler 
compounds. The zero energy corresponds to the Fermi level. Arrows depict the two spin directions.}
\label{fig2}
\end{figure*}

The rest twenty compounds under study are spin gapped materials as shown in 
Table\,\ref{table1}. The ones with less than 18 valence electrons per unit cell have 
negative total spin magnetic moments per unit cell and are p-type spin gapped metals. 
The compounds with more than 18 valence electrons, on the other hand, have positive 
total spin magnetic moments and are n-type spin gapped metals. All n-type (p-type) 
Spin gapped metals are not identical concerning their density of states. 
As shown in Table\,\ref{table1} there are several cases where the two spin band 
structures have a different character, \textit{i.e.}, the one spin channel presents 
either a normal metallic (NM) or semiconducting (SC) behavior while the other spin 
channel presents a n(p)-type spin gapped metallic behavior. To make this clear, in 
Figure\,\ref{fig2} we have gathered the total DOS for some selected compounds and 
compared them with the schematic representations in Figure\,\ref{fig1}. NiHfIn and 
NiHfSb are p-type and n-type gapped metals, respectively. FeTiSb presents p-type 
spin gapped metallic behavior for both spin directions. FeZrSb and FeVSn present p-type 
behavior for one spin direction and semiconducting (normal metallic) for the other. 
Similarly, PdTiSb, CoVSb and CuVSb present n-type Spin gapped metallic behavior for one 
spin direction and n-type/semiconducting/normal-metallic behavior for the other spin 
direction. None of the studied compounds presents an electronic band structure similar 
to the j panel in Figure\,\ref{fig1}, where one spin-band is n-type and the other p-type. 
One could envisage that this behavior can be identified in another class of Heusler 
compounds like the equiatomic quaternary Heuslers since several of them are type-II 
spin-gapless semiconductors \cite{gao2019high,Aull2019}. Actually in 
reference\,\cite{Ozdogan2013} (see figure 1), MnVTiAs compound combines a p-type gapped 
metallic behavior for the spin-up band structure with an n-type gapped metallic 
behavior for the spin-down electronic band structure. In this article, this behavior 
was referred to as "indirect spin-gapless semiconductors" since the concept of 
"Spin gapped metals" was not known.

In Table\,\ref{table1}, we have also included the atom-resolved spin magnetic moments for 
the two transition atoms in each compound, and the spin-polarization SP at the Fermi level. 
The atomic spin magnetic moments do not present any peculiarity. 
The transition metal atoms from Ti to Co are the ones carrying significant spin magnetic 
moments. On the other hand, the Ni/Cu atoms as well as the 4$d$ and 5$d$ transition metal 
atoms carry much smaller atomic spin magnetic moments. The spin-polarization SP at the 
Fermi level is defined as the difference between the number of spin-up 
$N^\uparrow(E_\mathrm{F})$ and spin-down $N^\downarrow(E_\mathrm{F})$ electronic states at 
the Fermi level divided by their sum, SP=$\frac{N^\uparrow(E_\mathrm{F})-
N^\downarrow(E_\mathrm{F})}{N^\uparrow(E_\mathrm{F})+N^\downarrow(E_\mathrm{F})}$. 
The SP is positive in all cases and reaches an ideal 100\%\ value when the spin-down 
electronic band is semiconducting.

As for the gapped metals, in Table\,\ref{table1} we also include the 
$W_{\mathrm{m}}^{\mathrm{I,E}(\uparrow/\downarrow)}$ and
$E_{\mathrm{g}}^{\mathrm{I,E}(\uparrow/\downarrow)}$ 
values for the spin gapped metals. The slash in each case separates 
the values corresponding to the spin-up and spin-down band structures. 
In the case of normal metallic or usual semiconducting behavior for one 
spin band structure, the former quantities cannot be defined and thus 
are not provided. Both $W_{\mathrm{m}}^{\mathrm{I,E}(\uparrow/\downarrow)}$ 
and $E_{\mathrm{g}}^{\mathrm{I,E}(\uparrow/\downarrow)}$ values 
range in almost all cases between a few tenths of an eV up to 1 eV.
It is worth noting that these value are very close to the energy gaps found 
in conventional  semiconductors like Si, Ge or GaAs where the width of the energy 
gap varies between $\sim$0.5 eV and $\sim$1.3 eV. Thus these spin gapped
metals can be combined  with conventional semiconductors in devices. 
Finally, the relative $W_{\mathrm{m}}^{\mathrm{I,E}(\uparrow/\downarrow)}$ and
$E_{\mathrm{g}}^{\mathrm{I,E}(\uparrow/\downarrow)}$ widths are material's specific, 
as it can be deduced from Table\,\ref{table1}, and no general rule can be drawn. 
This variety of values enhances the potential functionalities offered by the studied 
spin gapped metals.

Continuing our discussion on gapped metals and spin gapped metals, let's delve into their 
potential to revolutionize various electronic devices. These materials offer significant 
promise in addressing a major limitation of tunnel FETs and NDR tunnel diodes: band tails
tunneling caused by doped semiconductor electrodes. Unlike their doped counterparts, gapped 
metals exhibit electronic behavior akin to intrinsic semiconductors, effectively mitigating 
this undesirable effect. This unique property paves the way for the development of high-performance, 
steep-slope tunnel FETs. As a concrete example, we can envision a design utilizing a p-type 
17-valence electron half-Heusler compound as the source electrode, an 18-valence electron 
intrinsic semiconducting channel, and a 19-valence electron n-type  half-Heusler compound as 
the drain electrode. Moreover, gapped metals offer exciting possibilities for novel FET designs 
beyond tunnel FETs. By leveraging both p-type (17-valence electron) and intrinsic (18-valence 
electron) half-Heusler compounds as source-drain electrodes and channel material, a new class 
of FETs can be realized. These devices not only exhibit conventional transistor functionality 
but also possess the novel gate-voltage tunable NDR effect. The external band gap  
$E_{\mathrm{g}}^{\mathrm{E}}$ of  the p-type gapped metal, along with the bandwidth 
$W_{\mathrm{m}}^{\mathrm{E}}$, play a crucial role in determining the $I$-$V$ characteristics 
of the FET and manifestation of the NDR effect. Note that this behavior, previously predicted 
in FETs based on cold metal source-drain electrodes \cite{yin2022computational},  opens doors 
for innovative functionalities within a single device.

Finally, the potential of spin gapped metals extends far beyond the already impressive 
functionalities predicted for cold metals (or gapped metals) in devices like NDR tunnel 
diodes and steep-slope FETs \cite{wang2023cold}. However, the true power of these materials 
lies in their ability to exploit the spin degree of freedom, opening doors to entirely new 
functionalities. One exciting application would be replacing diluted magnetic semiconductors 
(DMS) in spin-Esaki diodes \cite{kohda2001spin,einwanger2009tunneling,kohda2006bias,anh2016observation,anh2018electrical,arakawa2020tunneling}. 
DMS are currently plagued by limitations such as low Curie temperatures and difficulties in 
achieving high spin polarization. Spin gapped metals, on the other hand, offer the potential 
to overcome these limitations due to their intrinsic magnetic properties and unique band 
structures. Furthermore, spin gapped metals hold promise for realizing a new class of steep-slope 
spin FETs that can exhibit both non-local giant magnetoresistance (GMR) and NDR effect. This 
unique combination could revolutionize magnetic memory and logic concepts, paving the way for 
advancements in logic-in-memory computing.

\section{Conclusions}

In conclusion, gapped metals, a recently introduced category of materials, offer exciting 
possibilities for nanoelectronic devices. Their unique band structure allows them to behave 
intrinsically as p- or n-type semiconductors, eliminating the need for external doping. 
Building upon this concept, we have introduced the concept of "spin gapped metals," which 
exhibit independent p- or n-type character for each spin channel. This intrinsic magnetism, 
analogous to diluted magnetic semiconductors, eliminates the need for transition metal doping. 
Unlike gapped metals (or cold metals), spin gapped metals open doors to entirely 
new functionalities, including next-generation spintronic devices like spin-Esaki diodes and
spin FETs with both non-local GMR effect and gate-voltage tunable NDR effect. To illustrate 
this concept, we employed first-principles electronic band structure calculations on half-Heusler 
compounds identified through the Open Quantum Materials Database. Our analysis, focusing on 
compounds with varying valence electron counts, revealed materials exhibiting both gapped 
metal and spin gapped metal behavior. This work highlights the transformative potential of 
spin gapped metals for materials design in next-generation spintronic and multifunctional 
devices. Further theoretical and experimental exploration is crucial to fully unlock the 
capabilities of this promising class of materials. Such studies could involve investigating 
their magnetic properties, transport characteristics, and device integration strategies.

\section*{Data Availability Statement}

Data available on request from the authors

\nocite{*}
\providecommand{\noopsort}[1]{}\providecommand{\singleletter}[1]{#1}%

\end{document}